\documentclass{ws-procs9x6}
\newcommand\bC{\mathbf{C}}

\newcommand\cO{\mathcal{O}}

\begin{document}

%%%%%%%%%%%%%%%%%%%%%%%%%%%%%%%%%%%%%%%%%%%%%%%%%%%%%%%%%%%%%% 
% title, author(s) and address(es) put here:                 %
%%%%%%%%%%%%%%%%%%%%%%%%%%%%%%%%%%%%%%%%%%%%%%%%%%%%%%%%%%%%%% 

\title{ B-branes and the derived category}

\author{Sheldon Katz}
\address{Departments of Mathematics and Physics\\
University of Illinois at Urbana-Champaign\\Urbana, IL 61801\\Email:
katz@math.uiuc.edu}


\maketitle

\abstracts{
By a direct CFT computation, the spectrum of the topological B-model
is compared to Ext groups of sheaves.  A match can only be made
if abstract vector bundles on holomorphic submanifolds
are twisted by the canonical $\mathrm{Spin}^c$
structure of its support in describing physical branes.
}

%%%%%%%%%%%%%%%%%%%%%%%%%%%%%%%%%%%%%%%%%%%%%%%%%%%%%%%%%%%%%
% The main text of your paper                               %
%%%%%%%%%%%%%%%%%%%%%%%%%%%%%%%%%%%%%%%%%%%%%%%%%%%%%%%%%%%%%

\section{Introduction.}

In recent years, there has  been much fruitful interplay between physics and
geometry through string theory.  Many questions in physics correspond
to questions in pure geometry.  The motivations in physics and
geometry are often of a completely different nature, so that finding
the precise dictionary between the two areas is key to making
fundamental progress.  
 
The spectrum of the closed topological string is well
understood\cite{witten}, and the correlation functions can be
computed.  In this talk, I outline a similar description for the open
string sector of the topological B-model on a compact K\"ahler
manifold $X$.  The first task is to understand the boundary
conditions, which are to be identified with the derived category in
well-known ways\cite{kont,ericdc,mike,al}.  The simplest case arises
from a holomorphic submanifold $i:S\hookrightarrow X$ and holomorphic
vector bundle $E$ on $X$, so that a D-brane is wrapped on $S$ with gauge
bundle $E$.  This data can be identified with the sheaf $i_*E$ of
$\cO_X$ modules, which is not a bundle unless $S=X$.  In any case,
$i_*E$ is an object of the derived category of $X$.  More general
objects of the derived category can be identified with boundary
conditions as well.  

This pleasing identification of geometric objects with boundary
conditions is complicated by worldsheet anomalies, which can be
eliminated by twisting the gauge bundle by a $\mathrm{Spin}^c$
structure\cite{fw}.  A canonical choice is given by the canonical
$\mathrm{Spin}^c$ structure on $X$.  This corresponds to modifying the
gauge bundle $E$ to become, somewhat imprecisely, $E\otimes
K_S^{-1/2}$.  With this choice, the open string spectrum matches Ext
groups exactly\cite{ks}.  With any other twisting or no twisting at
all, the spectrum can fail to match Ext groups.  So the identification
of objects of the derived category with branes is a bit more subtle
than one would like.

In this talk, I give the basic examples illustrating how to compute
the spectrum directly in BCFT and how the results are only sensible when
the anomaly is included in the analysis.  More details are available
elsewhere\cite{ks,ericlec}, as are generalizations to
orbifolds\cite{kps},
flat B-field backgrounds\cite{cks}, and 
D-branes wrapping nonreduced schemes\cite{dks,ericqts}.

\section{Closed string B model}  

I start by reviewing the topological $B$-model on a Calabi-Yau
threefold $X$.  Locally $X$ has three complex coordinates
$(\phi^1,\phi^2,\phi^3)$ which can be identified with bosonic fields on
the worldsheet.  A nowhere vanishing antisymmetric holomorphic tensor
$\Omega=\Omega_{ijk}$ is required to soak up fermion zero modes.  This
condition is satisfied for a Calabi-Yau threefold.  The topological
B-model can be compared to physical strings in 10D $M^4\times X$;
I do not do that here but make some brief comments at the end.

\smallskip
Tensors on $X$ have holomorphic indices $i$ and antiholomorphic
indices $\bar{i}$.  These are associated to the holomorphic and
antiholomorphic tangent bundles of $X$ which are global objects, i.e.\ 
\begin{eqnarray*}
\rho^i&\in T^{1,0}(X)\\
\rho^{\bar i}&\in T^{0,1}(X)
\end{eqnarray*}
with $\det(T^{1,0}(X))$ a trivial bundle, 
required by fermion zero modes coupling to $T^{1,0}(X)$ as mentioned above.

The worldsheet is a Riemann surface $\Sigma$ with local
complex coordinate $z$.  Covariant tensors split into
holomorphic and antiholomorphic pieces.
The holomorphic part is called the canonical bundle and is
denoted by $K_\Sigma=(T^{1,0}(\Sigma))^*$.
The antiholomorphic part is denoted by $\bar{K}_\Sigma=(T^{0,1}(\Sigma))^*$

The fields in the topological string combine the tensor structures on 
$\Sigma$ and $X$.  So the fields in the theory are:

\bigskip
\begin{eqnarray*}
&\ \phantom{xxx}\phi^i(z) \\
\psi_{\pm}^{\overline{\imath}}(z) & \in & 
\Gamma\left( \phi^* T^{0,1} X \right), \\
\psi_+^i(z) & \in & \Gamma \left( K \otimes
\phi^* T^{1,0} X \right), \\
\psi_-^i(z) & \in & \Gamma\left( \overline{K} \otimes
\phi^* T^{1,0} X \right)
\end{eqnarray*}

The $\psi$ fields anticommute with each other, and the $\phi$
are bosonic.  The bulk B model Lagrangian is
\begin{eqnarray*}
&\frac{1}{2} g_{i \overline{\jmath} } \partial \phi^{i}  
\overline{\partial} \phi^{ \overline{\jmath} } \: + \:
\frac{1}{2} g_{i \overline{\jmath} } \partial \phi^{ \overline{\jmath} }
\overline{\partial} \phi^{i} \: + \:
i g_{i \overline{\jmath} } \psi_{-}^{ \overline{\jmath} }
D_z \psi_{-}^{i} \: + \: \\
&
i g_{ i \overline{\jmath} } \psi_{+}^{\overline{\jmath}} 
D_{ \overline{z} } \psi_{+}^i \: + \:
R_{i \overline{\imath} j \overline{\jmath} }
\psi_+^i \psi_+^{\overline{\imath}}
\psi_-^j \psi_-^{\overline{\jmath}}
\end{eqnarray*}
with BRST transformations
\begin{eqnarray*}
\delta \phi^i & = & 0, \\
\delta \phi^{ \overline{\imath} } & = & i \alpha \left(
\psi_+^{\overline{\imath}} \: + \: \psi_-^{ \overline{\imath} } \right), \\
\delta \psi_+^i & = & - \alpha \partial \phi^i, \\
\delta \psi_+^{\overline{\imath}} & = & -i \alpha
\psi_-^{ \overline{\jmath}} \Gamma^{\overline{\imath}}_{ \overline{\jmath}
\overline{m} } \psi_+^{ \overline{m} }, \\
\delta \psi_-^i & = & - \alpha \overline{\partial} \phi^i, \\
\delta \psi_-^{ \overline{\imath} } & = & - i \alpha
\psi_+^{ \overline{\jmath} } \Gamma^{ \overline{\imath} }_{
\overline{\jmath} \overline{m} } \psi_-^{ \overline{m} }.
\end{eqnarray*}

The BRST symmetry satisfies $\delta^2=0$.  Consideration is restricted 
to BRST-closed operators, $\delta F = 0$.
Addition of a BRST-exact operator does not alter correlation functions
\[
F\sim F+\delta G,
\]
so the spectrum is described by BRST cohomology
\[
H^*_{\mathrm{BRST}}=\frac{\left\{{\rm BRST-closed\ operators}\right\}}
{\left\{{\rm BRST-exact\ operators}\right\}}.
\]

It is convenient to change variables and 
define
\begin{eqnarray*}
\eta^{ \overline{\imath} } & = & \psi_+^{ \overline{\imath} } \: + \:
\psi_-^{ \overline{\imath} }, \\
\theta_i & = & g_{ i \overline{ \jmath} } \, \left(
\psi_+^{ \overline{\jmath}} \: - \: \psi_-^{\overline{\jmath}}
\right), \\
\rho_z^i & = & \psi_+^i, \\
\rho_{ \overline{z} }^i & = & \psi_-^i,
\end{eqnarray*}

A vertex operator is of the form
\[
b^{j_1 \cdots j_m}_{\overline{\imath}_1 \cdots 
\overline{\imath}_n}(\phi_0) \, \eta^{ \overline{\imath}_1} \cdots
\eta^{ \overline{\imath}_n} \theta_{j_1} \cdots \theta_{j_m}
\]
This operator can be identified with a tensor field on $X$:
\[
b^{j_1 \cdots j_m}_{\overline{\imath}_1 \cdots 
\overline{\imath}_n}(\phi)d\bar{\phi^{i_1}}\wedge\ldots\wedge
d\bar{\phi^{{i_n}}}\otimes \frac{\partial}{\partial \phi^{j_1}}\wedge\ldots\wedge
\frac{\partial}{\partial \phi^{j_m}}
\]
where $\wedge$ denotes a completely antisymmetrized product.

Under this identification, BRST-cohomology is identified with the
Dolbeault cohomology of $\Lambda T^{1,0}(X)$, an important geometric
invariant of the holomorphic structure of $X$.

\[
H^q\left(\Lambda^pT^{1,0}(X)\right)=
\frac{\left\{\bar\partial-{\rm closed\ }(0,q)\ {\rm forms}\right\}}
{\left\{\bar\partial-{\rm exact\ }(0,q)\ {\rm forms}\right\}}
\]
where the forms take values in $\Lambda^pT^{1,0}(X)$.
Here $\bar\partial$ is the antiholomorphic differential, e.g.

\[
\bar\partial f = \frac{\partial f}{\partial \bar{\phi^i}}d\bar{\phi^i}.
\]
Among these Dolbeault cohomology classes, $H^1(T^{1,0}(X))$ parametrizes the holomorphic
structure.

Correlation functions of the associated vertex operators are
important.  For example, three point correlators of operators
associated to $H^1(T^{1,0}(X))$

\[
\langle A^{i_1}_{\bar{j}_1}\theta_{i_1}\eta^{\bar{j}_1},
 A^{i_2}_{\bar{j}_2}\theta_{i_2}\eta^{\bar{j}_2},
 A^{i_3}_{\bar{j}_3}\theta_{i_3}\eta^{\bar{j}_3} \rangle
\]
are identified with the Yukawa couplings in the gauge sector of the
heterotic string.

This is the kind of description to be extended to the open string.

\section{Open string B-model}

In extending the previous analysis to open string the first step
is to find the boundary conditions.
The simplest supersymmetric B-branes are described by a holomorphic
submanifold $S\subset X$ together with a gauge field $A_i$ on $S$.
The coordinates on $S$ are Neumann directions while coordinates
transverse to $S$ are Dirichlet directions.  Using the gauge field, 
introduce the usual boundary term in the action
\[
\int_{\partial\Sigma}A_i\frac{\partial\phi^i}{\partial t}+\ldots.
\]

\noindent

Along Neumann directions, the boundary conditions are
\[
\psi_+^i|_{ \partial \Sigma} \: = \: \psi_-^i |_{ \partial \Sigma},\qquad
\theta_i = \left( \mbox{Tr } F_{ i \overline{\jmath}} 
\right) \eta^{
\overline{\jmath}},
\]
$F$ being the field strength, while along Dirichlet directions, the boundary
conditions are

\[
\psi_+^i |_{ \partial \Sigma} \: = \:
- \psi_-^i |_{ \partial \Sigma},\qquad
\eta^{\overline{\imath}} = 0.
\]
Some care is needed in interpreting the intrinsic meaning of a Dirichlet
direction, since the tangent bundle of $X$ along $S$ has a $C^\infty$
splitting into the sum of the tangent bundle to $S$ and the normal
bundle to $S$ in $X$, i.e.\ $TX|_S\simeq TS\oplus N_{S/X}$, but there need
not be a holomorphic splitting.

The vertex operators are

\begin{displaymath}
b^{\alpha \beta j_1 \cdots j_m}_{\overline{\imath}_1 \cdots 
\overline{\imath}_n}(\phi_0) \, \eta^{ \overline{\imath}_1} \cdots
\eta^{ \overline{\imath}_n} \theta_{j_1} \cdots \theta_{j_m}
\end{displaymath}
where the $\bar{i}$ are indices on $S$ and the $\theta_j$ are constrained
by the boundary conditions.  Also,
$\alpha$, $\beta$ are gauge degrees of freedom (Chan-Paton indices).

The immediate goal is to evaluate the BRST cohomology and identify it
with a geometric object.  Mathematically, a novel complex of differential
forms has been defined, the novelty being the boundary conditions, 
and the problem is to compute its cohomology.
The answer is expected to match Ext
groups of objects in the derived category of algebraic geometry.
This does check, with some surprises along the way.

Now
$U(n)$ gauge fields live in rank $n$ vector bundles $E$ on $S$. Locally on
$S$, $E$ takes the form $U_{\alpha}\times\bC^n$ for $U_\alpha\subset S$.
Vector bundles are glued together over intersections $U_\alpha\cap
U_\beta$ via $r\times r$
      matrices $g_{\alpha\beta}$ acting on $\bC^r$ by multiplication.
Sections of bundles are identified with
local $\bC^r$-valued functions $s_\alpha$ on each $U_\alpha$
satisfying 
\begin{equation}
\label{transition}
s_\alpha=g_{\alpha\beta}s_\beta.
\end{equation}
The sections of $E$ form an important object of geometry called
a {\em coherent sheaf\/} on $X$.  This means that
sections $s_\alpha$
of $E$ can be multiplied by holomorphic functions $f$ on $X$, and that 
the result only depends on the restriction of such functions to $S$
(coherence is a technical finiteness condition which is automatically
satisfied here).    
Explicitly the multiplication is simply the identity 
$f|_S\cdot s_\alpha=g_{\alpha\beta}((f|_S)\cdot s_\beta)$ 
obtained from (\ref{transition})
after multiplication by $f|_S$.
The resulting sheaf is denoted $i_*E$ where $i:S\hookrightarrow
X$ is the inclusion.

Coherent sheaves form the heart of a t-structure on the derived
category of $X$.  From this heart, all objects of the derived category
can be built by shifts and extensions.  The heart, shifts, and
extensions model branes, antibranes, and tachyon condensation.  The
notion of stability of bundles has a corresponding physical
interpretation ({\em pi-stability})\cite{mike} which is being
developed mathematically as well.\cite{bridge,gkr}.  The surprises
which are described in this talk will need to be incorporated into
these notions.

\smallskip
Naively, the brane described by $E$ is to be identified with the sheaf
$i_*E$ of the derived category.  But this is not quite correct due to
anomalies; the standard correspondence between objects of the derived
category and branes has to be modified.

Now if two branes $S,\ T$ are considered, with strings from $S$ to $T$,
there are various combinations of Dirichlet and Neumann conditions at
the respective ends of the string.
The fermion zero modes couple to 
directions in $S$ and $T$ as well as to 
directions not in either $S$ nor $T$.
Geometrically, these correspond to
\[
T(S\cap T),\qquad \mathcal{N}=\frac{T(X)}{T(S)\oplus T(T)}
\]
respectively, where the
notational simplification of writing e.g.\ $T(S)$ in place of $T^{1,0}(S)$
has been adopted.
The triviality of
\[
\det(T(S\cap T))\otimes \det(\mathcal{N})
\]
is required to soak up fermion zero modes.
This is equivalent to the triviality of the bundle
\[
\det\left(N_{S\cap T,T}\right)\otimes \det\left(N_{S\cap T,S}\right),
\]
where 
$N_{S\cap T,T}$ is the bundle of normal vector fields to $S\cap T$ in $T$.

\section{The Freed-Witten anomaly}

An important anomaly in the path integral measure was described
by Freed and Witten\cite{fw}.
There are two interesting factors in the path integral:
\[
\mathrm{Pfaff}(D),
\]
the Pfaffian of the worldsheet Dirac operator, and
\[
\mathrm{exp}\left(i\int_{\partial\Sigma}A\right)
\]
They found that the Pfaffian has an anomaly,
characterized by the second Stiefel-Whitney class $w_2(S)\in
H^2(S,\mathbf{Z}_2)$: that in traversing a family of images of the
boundary of the worldsheet $\partial \Sigma$ parametrized by a curve
$C$,
\[
\mathrm{Pfaff}(D)\mapsto (-1)^{\langle w_2(S),C\times\partial\Sigma \rangle}
\mathrm{Pfaff}(D).
\]
So $\mathrm{exp}\left(i\int_{\partial\Sigma}A\right)$ must transform in the
same way to cancel the anomaly:
\[
\mathrm{exp}\left(i\int_{\partial\Sigma}A\right)\mapsto
(-1)^{\langle w_2(S),C\times\partial\Sigma \rangle}
\mathrm{exp}\left(i\int_{\partial\Sigma}A\right)
\]
This says that the gauge field does {\em not\/} live in a bundle, but
rather in a {\em twisted\/} bundle.  In fact, the gauge field must live
in a twisted bundle 
$E$ of the form
\[
E = B \otimes \left(L\right)^{1/2},
\]
where $B$ is an ordinary bundle, and $L$ is a line bundle with 
\begin{equation}
\label{w2}
c_1(L)\equiv w_2(S) \ (\mathrm{mod}\ 2).
\end{equation}

This suggests that the map 
\[
{\rm Bundles\ on\ }S\to\ {\rm physical\ branes}
\]
should not be the identity map, but should be modified to 
\[
E\mapsto E\otimes L^{1/2}.
\]
for some $L$.  The line bundle $L$ corresponds to a choice of $\mathrm{Spin}^c$
structure on $X$.  Any bundle $L$ satisfying (\ref{w2}) will give
an identification, but only one choice will identify the open string spectrum
with Ext groups of sheaves.

A complex manifold $X$ need not be spin, but is always
$\mathrm{Spin}^c$.  In fact, there is a canonical $\mathrm{Spin}^c$
structure\cite{lm}, and it is not a surprise that this is the one that
gives the twisting of the gauge bundle that relates the spectrum to
Ext groups.  The result is that the correct identification of bundles
with twisted sheaves is given by
\[
E\mapsto E\otimes\left(K_S^*\right)^{1/2}.
\]
Said differently, this corresponds to a trivialization of a twisting
of K-theory dictated by the anomaly.

\section{Computation of the spectrum}

With all the machinery in place, the method for computing
the spectrum can now be outlined in the case of
parallel coincident branes on $S \hookrightarrow X$.
Denote by $E,F$ the gauge bundles on the respective ends of the string.
The boundary vertex operators are of the form
\begin{displaymath}
b^{\alpha \beta j_1 \cdots j_m}_{\overline{\imath}_1 \cdots 
\overline{\imath}_n}(\phi_0) \, \eta^{ \overline{\imath}_1} \cdots
\eta^{ \overline{\imath}_n} \theta_{j_1} \cdots \theta_{j_m}.
\end{displaymath}
 
Recall the effect of the boundary conditions: the
$\theta$ indices are determined by their values in
directions normal to $S$, 
the $\eta$ indices are constrained
to only live along directions tangent to $S$, and
$\phi$ zero modes
$\phi_0$ are constrained to only map out $S$.

By explicit computation\cite{ks}, the BRST cohomology classes are in one-to-one
correspondence with ext groups
\[
\mathrm{Ext}_X^n(i_*(E),i_*F)
\]
In particular, if $E=F$ then 
$\mathrm{Ext}_X^1(i_*E,i_*E)$ describes first order
deformations of $E$ as a sheaf or object of the derived category.

Here is a small part of the computation.
Restricting attention to $\theta$ indices tangent to $S$, the BRST condition
determines
bundle-valued differential forms living on $S$, i.e.\ 
\[
b^{\alpha \beta j_1 \cdots j_m}_{\overline{\imath}_1 \cdots 
\overline{\imath}_n}(\phi_0) \, \eta^{ \overline{\imath}_1} \cdots
\eta^{ \overline{\imath}_n} \theta_{j_1} \cdots \theta_{j_m}
\in H^n \left(S, E^* \otimes F \otimes 
\Lambda^m N_{S/X} \right)
\]
where $N_{S/X}$ is the normal bundle to $S$ in $X$.

Putting the boundary conditions
\[
\theta_i = \left( \mbox{Tr } F_{ i \overline{\jmath}} 
\right) \eta^{
\overline{\jmath}}
\]
back in iteratively, the {\em local to global spectral sequence}
\[
\label{easyss}
E_2^{p,q}:
H^p \left(S, \mathcal{E}^{\vee} \otimes \mathcal{F} \otimes \Lambda^q N_{S/X} 
\right) \: \Longrightarrow \:
\mbox{Ext}^{p+q}_X\left( i_* \mathcal{E}, i_* \mathcal{F} \right)
\]
is reproduced exactly, each iteration corresponding to an iteration of
the spectral sequence.  This shows that the BRST cohomology is the abutment
of the spectral sequence, i.e.\ the anticipated groups
$\mbox{Ext}^{n}_X\left( i_* \mathcal{E}, i_* \mathcal{F} \right)$.

The Freed-Witten anomaly is not manifest here --- it cancels out of
the opposite ends of the string, so has been suppressed from the
notation.  But this is not the case in general.  Consider branes 
\[
i:S\hookrightarrow X,\ {\rm gauge\ field\ in\ } E\otimes(K_S^*)^{1/2}
\]
\[
j:T\hookrightarrow X,\ {\rm gauge\ field\ in\ } F\otimes(K_T^*)^{1/2}
\]
Now the
BRST cohomology is on one-to-one correspondence with elements of
\[
\mathrm{Ext}^n_X(i_*E,j_*F))
\]
The 
computation matches the local to global spectral sequence with $E_2$ term

\[
E_2^{p,q} \: = \: H^p\left(S\cap T, E^*|_{S \cap T}
\otimes F|_{S \cap T} \otimes \Lambda^{q-m} \mathcal{N}
\otimes \det N_{S \cap T / T} \right)
\]
\[
\ \qquad \Longrightarrow 
\mbox{Ext}^{p+q}_X \left( i_* E, j_* F \right) 
\]
where $m$ is the rank of $N_{S \cap T/T}$ and
\[
\mathcal{N}=\frac{T(X)}{T(S)\oplus T(T)}.
\]
as before.  Note that the term $\det N_{S \cap T / T}$ is physically
absent if the anomaly is ignored.  The computation works only because
the B-model anomaly cancels the Freed-Witten anomaly.  To see this, use the
identity
\[
\sqrt{ \frac{K_S|_{S\cap T}}{K_T|_{S\cap T}} }  \cong 
\det N_{S \cap T/T} 
\]

This may seem a bit odd that the B-model anomaly is required, since
for physical strings, the condition on the triviality of
\begin{equation}
\label{triv}
\det\left(N_{S\cap T,T}\right)\otimes \det\left(N_{S\cap T,S}\right)
\end{equation}
is absent, while there is general confidence that
the derived category and ext groups gives the correct answer.
This suggests that for physical strings (\ref{triv}) is a new
supersymmetry selection rule.

I would like to use this description of the spectrum to compute
correlation functions, but there are a number of serious obstacles.
All that can be said at present is that 
three-point functions can be computed via the Yoneda pairing.

\section{Generalizations}

The analysis above extends to a wide range of string models, with
similar results.  The extension to orbifolds\cite{kps} shows that this
analysis is not constrained to large radius and shows how Ext groups
in (quotient) stacks arise in string theory.  The extension to flat
$B$-field backgrounds\cite{cks} shows how Ext groups in the derived
category of twisted sheaves arises in string theory.  Finally, the
boundary conditions can correspond to sheaves on nonreduced
schemes, which can sometimes be described physically by turning on
nilpotent Higgs fields.  The resulting spectrum coincides with
Ext groups of sheaves associated to these nonreduced
schemes\cite{dks}.

%%%%%%%%%%%%%%%%%%%%%%%%%%%%%%%%%%%%%%%%%%%%%%%%%%%%%%%%%%%%%
%                                                           %
% You may repeat \section{SECTION N-th HEADING TYPE HERE}   %
%                                                           %
% Do start a subsection or sub-subsection, do this:-        %
%                                                           %
%   \subsection{SUBSECTION HEADING TYPE HERE}               %
%                                                           %
%   \subsubsection{SUBSUBSECTION HEADING TYPE HERE}         %
%                                                           %
% instead of the above                                      %
%                                                           %
%%%%%%%%%%%%%%%%%%%%%%%%%%%%%%%%%%%%%%%%%%%%%%%%%%%%%%%%%%%%%

\section{CONCLUSIONS}

Open string vertex operators can be explicitly and completely 
described by complex geometry, in the situation where the
boundary conditions correspond to
gauge fields in (twisted) bundles on holomorphic submanifolds
and beyond.  It is desirable to
describe the twisting needed to cancel the
anomaly intrinsically in terms
of the derived category.
Thus boundary branes can be described directly in CFT.  
It is desirable to extend existing methods to 
compute higher correlation functions.

%%%%%%%%%%%%%%%%%%%%%%%%%%%%%%%%%%%%%%%%%%%%%%%%%%%%%%%%%%%%%
% Doing Acknowledgement                                     %
%%%%%%%%%%%%%%%%%%%%%%%%%%%%%%%%%%%%%%%%%%%%%%%%%%%%%%%%%%%%%

\section*{Acknowledgments}

I would like to thank Eric Sharpe for many valuable discussions and
for collaboration on this topic, and Robert Karp for comments on
an earlier draft.

%%%%%%%%%%%%%%%%%%%%%%%%%%%%%%%%%%%%%%%%%%%%%%%%%%%%%%%%%%%%%
% Doing references:                                         %
%%%%%%%%%%%%%%%%%%%%%%%%%%%%%%%%%%%%%%%%%%%%%%%%%%%%%%%%%%%%%

\end{document}